\documentclass[useAMS,usenatbib]{mn2e}
\usepackage{amssymb}
\usepackage{amsmath}
\usepackage{txfonts}
\usepackage{graphicx}
\usepackage{natbib}
\usepackage{rotating}
\usepackage{url}
\usepackage{epsfig}
 
\newif\ifpdf\ifx\pdfoutput\undefined\pdffalse\else\pdfoutput=1\pdftrue\fi

\def\intgr{\textit {INTEGRAL}}

\def\rxte{\textit {RXTE}}
\def\asca{\textit {ASCA}}

\def\xmm{\textit {XMM-Newton}}
\def\xmmsp{\textit {XMM-Newton }}

\def\qso {QSO 0241+622}
\def\deg {$^\circ$}
\def\lsi {LSI~+61\deg~303}

\usepackage{color}
\definecolor{red}{rgb}{0.7,0,0}
\definecolor{blue}{rgb}{0,0,0.7}

\begin{document}

\title{\intgr\ and \xmm\ observations of  \lsi} 

 \author[M. Chernyakova et al.]{M. Chernyakova$^{1,2}$\thanks{E-mail:Masha.Chernyakova@obs.unige.ch}\thanks{M.Chernyakova 
is on leave from Astro Space Center of the P.N.~Lebedev Physical
Institute,  Moscow, Russia}, A. Neronov$^{1,2}$, R. Walter$^{1,2}$\\
$^{1}$INTEGRAL Science Data Center, Chemin d'\'Ecogia 16, 1290 Versoix, Switzerland\\
$^{2}$Geneva Observatory, 51 ch. des Maillettes, CH-1290 Sauverny, Switzerland}

\date{Received $<$date$>$  ; in original form  $<$date$>$ }

\pagerange{\pageref{firstpage}--\pageref{lastpage}} \pubyear{2005}

\maketitle

\label{firstpage}

\begin{abstract} 
\lsi\ is one of the  few X-ray binaries with Be star companion  from which 
both radio  and high-energy gamma-ray emission  have been observed.  We present
\xmm\ and \intgr\ observations which reveal variability of the X-ray spectral
index of the system. The X-ray spectrum is hard (photon index $\Gamma\simeq
1.5$) during the orbital phases of both high and low X-ray flux. However, the
spectrum softens at the moment of  transition from high to low X-ray state. The
spectrum of the system in the hard X-ray band  does not reveal the presence of
a cut-off (or, at least a spectral break)  at 10-60~keV energies,  expected if
the compact object is an accreting neutron star.  The observed spectrum and
spectral variability can be explained if the compact object in the system is a
rotation powered pulsar. 
\end{abstract}


\section{Introduction}

The Be star binary \lsi\ is one of the few X-ray binaries (along with PSR
B1259-63 and LS 5039) from which radio  and very high-energy gamma-ray emission is
observed. Radio emission from the system is highly variable and shows
periodicity of $T=26.4960$ days, which can be associated with the binary
orbital period \citep{gregory02}. Radio observations reveal the presence of
100~AU-scale jet in the system which places \lsi\ among the Galactic
micro-quasars \citep{massi93,massi04}. The system is also a Galactic
"micro-blazar"  due to its  association  with 100 MeV gamma-ray source 2CG
135+01 \citep{tavani98} visible up to TeV energies \citep{albert06}. 

\citet{mendelson89} found an optical modulation in the V-band with the same
periodicity as the radio periodicity. This is confirmed by Paredes et al.
(1994); the optical modulation in the V-band is about 0.15 mag.  Optical data
allow to constrain the orbital parameters of the system revealing the
eccentricity of the orbit, $e\simeq 0.7$ \citep{casares05}. However, the
measurements are not sufficient to determine the nature of the  compact object
(neutron star or black hole), because the inclination of the orbit is poorly
constrained.  

The X-ray emission from \lsi\ is also variable  and has nonthermal spectrum.
Since the discovery, the system was twice monitored simultaneously in the 
X-ray and radio bands over a single orbital cycle \citep{taylor96,harrison00}. 
These simultaneous observations  show that X-ray emission peaks almost half an
orbit before the radio. In both X-ray and radio bands the orbital phases of
maxima of the flux "drift" from orbit to  orbit. 

Two major types of models of radio-to-X-ray activity of \lsi\ were proposed in
the literature. Models of the first type, first introduced by \citet{taylor84}
assume that activity of the source is powered by accretion onto the compact
object. In the second class of models, first proposed by \citet{maraschi81},
the activity of the source is explained by interactions of a young rotation
powered pulsar with the wind from the companion Be star. 

If the system is an accreting neutron star or black hole, one expects to find a
cut-off powerlaw spectrum in the hard X-ray band.   The cut-off energy is
normally at $10-60$~keV \citep{white83,filippova05} for neutron stars and at
$\sim 100$~keV for  black holes \citep{mcclintock03} (the cut-off can move to even 
higher energies in the "high/soft" state see e.g. \citet{belloni}).   Of course, it is
possible that the emission from the "central engine" of the  micro-blazar \lsi\
is  "masked" by the emission from the jet, which can be beamed toward an
observer on Earth,  similarly to what is observed in blazars.   In this case
the hard X-ray spectrum of the source is a superposition of the powerlaw  jet
emission and emission from the accretion disk. If the jet and accretion
contributions to the X-ray spectrum are comparable, then emission from the
accretion disk should at least produce an observable  spectral feature (e.g. a 
bump, a break or turnover) in the  10-100~keV energy band. 

Below we present a study of \lsi\ in the 0.5-100~keV  energy band with \xmm\
and \intgr\ which shows that the spectrum of the source is well fit by a simple
powerlaw, without any signature of high-energy cut-off. Moreover, the X-ray
powerlaw matches smoothly to the higher-energy spectrum in the 100~keV --
10~MeV band found with the {\it CGRO} instruments OSSE \citep{tavani96} and
COMPTEL \citep{vandijk96}. The broad band spectrum of the source does not exhibit
a cut off up to at least GeV energies. The spectrum of \lsi\ in the X-ray -- soft
gamma-ray band is thus very different from the typical  spectra of  accreting
neutron stars and black holes.

Contrary to accreting compact object models, a  featureless powerlaw  
keV~--~MeV spectrum is expected in the "rotation powered pulsar" scenario.
Unfortunately, the population of X-ray binaries with a rotation powered pulsars
as compact objects is not so rich as that of accreting pulsars. In fact, the
only firmly established example of X-ray binary with a young radio pulsar as a
compact object, PSR B1259-63 \citep{johnston92}, exhibits a featureless
powerlaw spectrum in the X-ray band which smoothly continues to the higher
energies, exactly as in \lsi. Moreover, the overall spectral energy
distributions and variability properties of the two systems are qualitatively
very similar: large radio outbursts which occur once per orbit, similar X-ray
spectra and high-energy gamma-ray emission  extending up to the TeV energy
band. Recent observations of PSR B1259-63 during its 2004 periastron passage 
\citep{chernyakova06} show that the X-ray emission from the system  is most
probably produced via   inverse Compton scattering of the UV radiation from the
Be star by the pulsar wind electrons, the same mechanism  as proposed by
\citet{maraschi81}  for  \lsi

The \xmm\ and \intgr\ observations of \lsi\, reported here,  reveal the
variability of the X-ray spectrum of the system during the transition from high
to low X-ray state. We find that the spectra of the system in high and low
X-ray states are very similar, but during the transition to the low flux state
the spectrum softens. It turns out that explanation of such behavior of the
system is not trivial and, in fact, the shape and variability of the spectrum
put tight constraints on the injection mechanism of high-energy electrons
responsible for the X-ray emission as well as on the geometry of the X-ray
emission region.

This paper is organized as follows: in Section 2 we describe the details of the
\intgr\ and \xmmsp  data analysis.  The results (imaging, spectral and timing
analysis)  are presented in Section 3. In Section 4 we discuss the physical
implications of the results for the synchrotron -- inverse Compton broad band
model of the source. This gives us the possibility to constrain physical
parameters of the system and develop a physical model of the source which we
study in more details in Section 5.  Finally, we summarize our findings in
Section 6.  

\section[]{Observations and Data Analysis}

\subsection{\intgr\ observations}

Since the launch of \intgr\ \citep{winkler} on  October 17, 2002, \lsi\ was
several times in the field of view of the main instruments   during the routine
Galactic plane scans and pointed observations  (see Table \ref{intdata} for
details).  Most of the times the distance of the source  from the center of the
field of view  was too large to use  the X-ray monitor JEM-X.  Therefore IBIS/ISGRI
\citep{lebrun} is the only instrument we can use in our analysis of this
source. In this analysis we have used the version 5.1 of the Offline Science
Analysis (OSA) software distributed by the ISDC \citep{courvoisier03}.

In our analysis we have used all available public data spread over the period
from the January 2003 (rev 25)  until March 2005 (rev 288). Overall we have 
analyzed 600 science windows which resulted in an effective vignetting
corrected exposure of 273~ksec.

In order to study the spectral variations of the source we have grouped the
data into three parts covering orbital phases 0.4 -- 0.6, 0.6 -- 0.8 and  0.8
-- 0.4 (see Table \ref{intdata}). The spectral analysis was done with the use of the 
\texttt{mosaic\_spec} tool, which  extracts spectra for a given sky position from 
mosaic sky images.

\begin{table}
\caption{Journal of the \intgr\  observations of \lsi. $^*$ \label{intdata}}
\begin{tabular}{c|c|c|c|c|c}
\hline
 Data & Orbital phase &  Effective   &20-60 keV Flux&$\Gamma$ \\
  Set &               &     Exposure  (ks)  &$10^{-11}$ergs s$^{-1}$cm$^{-2}$\\
\hline
I1  & 0.4 -- 0.6  &50& 3.8$\pm 0.6$ &1.7$\pm 0.4$ \\ 
I2  & 0.6 -- 0.8  &23& 3.0$\pm 1.0$&3.6$^{+1.6}_{-1.1}$ \\
I3  & 0.8 -- 0.4  & 200&2.4$\pm0.3$ &1.4$\pm 0.3$ \\
I$_{\mbox{av}}$& &273&2.5$\pm 0.3$ &1.6$\pm 0.2$ \\
\hline
\end{tabular}
$^*$ Given errors represent 68\% confidence interval 
uncertainties.
\end{table}

\subsection{\xmm\ observations}

\xmm\ has observed \lsi\  with the EPIC instruments  five times during 2002.
Four observations have been done during the same orbital cycle, and the fifth
one  has been done seven months later.					    
The log of the \xmm\ data analyzed in this paper is presented in
Table~\ref{data}.

\begin{table}
\caption{Journal of \xmm\ observations of \lsi \label{data}}
\begin{tabular}{c@{\,}c@{\,}ccc@{\,}c}
\hline
Data& Observational& Date &    MJD&  Orbital&Exposure (ks)\\
 Set&   ID         &      & (days)&    Phase  &MOS1/MOS2/PN     \\
\hline
X1& 0112430101& 2002-02-05&52310  & 0.55  &5.8/5.8/5.0  \\
X2& 0112430102& 2002-02-10&52315  & 0.76  &5.8/5.8/5.0  \\
X3& 0112430103& 2002-02-17&52322  & 0.01  &1.5/1.5/5.0  \\
X4& 0112430201& 2002-02-21&52326  & 0.18  &7.0/7.0/5.1  \\
X5& 0112430401& 2002-09-16&52533  & 0.97  &6.2/6.2/6.0  \\
\hline
\end{tabular}
\end{table}

The \xmmsp Observation Data Files (ODFs) were obtained from the on-line Science
Archive\footnote{http://xmm.vilspa.esa.es/external/xmm\_data\_acc/xsa/index.shtml};
the data were then processed and the event-lists filtered using {\sc xmmselect}
within the Science Analysis Software ({\sc sas}) v6.0.1. In all observations
the source was observed with the PN detector in the Small Window Mode, and with
MOS1 and MOS2 detectors in the Full Frame Mode.  In all observations a medium
filter was used.

The event lists for spectral analysis were extracted from a
40$^\prime$$^\prime$ radius circle at the source position for  MOS1,  from a
30.0$^\prime$$^\prime$ radius circle for MOS2, and for PN  a region of
25$^\prime$$^\prime$ around the source position was chosen.  Background photons
were collected from a region located in the vicinity of the source with the
same size area as the one chosen for the source photons.

For the spectral analysis, periods affected by soft proton flares need to be
filtered out. To exclude them we have used script \texttt{xmmlight\_clean.csh}
\footnote{http://www.sr.bham.ac.uk/xmm2/xmmlight\_clean.csh}. The source
countrate in 2  -- 10 keV energy range  varied from about 0.03  to 0.06 cts/s
for MOS  instruments, and from 0.05 to 0.10 cts/s for PN.
With the help of the \texttt{GRPPHA} FTOOL we have re-binned the spectra, 
so that  each energy bin has at least 30 counts.

Data from MOS1, MOS2 and PN detectors were combined in the spectral analysis to
achieve better statistics.

\section{Results}

\subsection{Imaging Analysis}

\begin{figure}
\begin{center}
\includegraphics[width=8cm,angle=0]{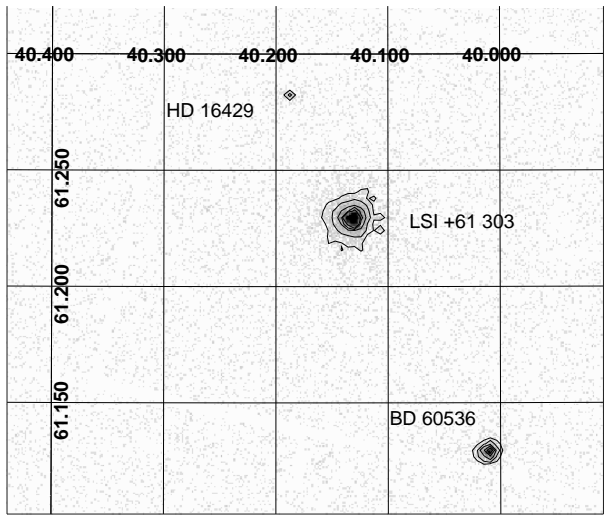}
\end{center}
\caption{
Contour plot of the \xmm\ field of view for the X1 observation. A total of 5
contours were used with a square root scale between  1 counts per pixel (outer
contour) and 25 counts per pixel.}
\label{xmm_ima}
\end{figure}

Three sources were clearly detected in the \xmm\ field of view during its \lsi\
observation  program  (see Fig. \ref{xmm_ima} for the contour plot of \xmm\
field of view for the MJD 52310 (X1) observation). Besides \lsi\ itself we
have  detected  two more sources that were also detected by ROSAT, and were
identified as two optical sources BD +60 536, and HD +16 429 by \citet{GM95}.
In the X-rays  BD +60 536 has also been  detected by \textit{ASCA} in 1994
\citep{leahy97}. HD +16 429 has also been  observed in radio  during deep VLA
observations of the \lsi\ field \citep{marti98}.

\begin{figure}
\begin{center}
\includegraphics[width=8.5cm,angle=0]{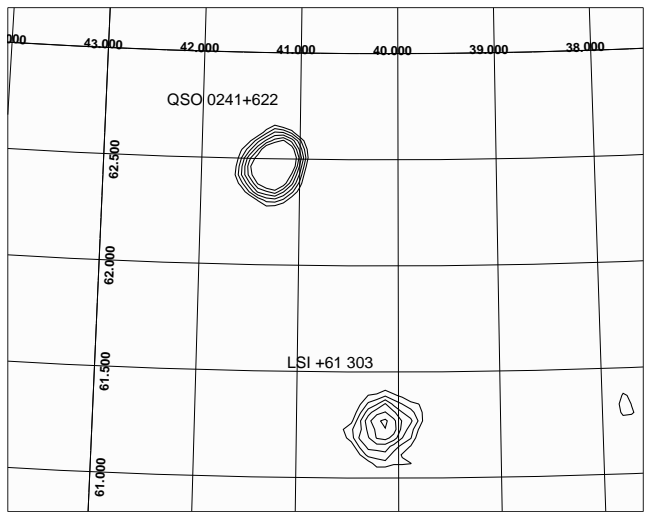}
\end{center}
\caption{Contour plot of the \intgr\ significance map of the \lsi\
region in  22.1-63.2 keV energy range with a 273 ks effective exposure. 
A total of 6 contours were used with a linear scale between 3 $\sigma$ (outer contour) 
and 8 $\sigma$ (inner contour).}
\label{int_ima}
\end{figure}

The presence of a quasar \qso\ some $1.4^\circ$ away from \lsi\ was a problem
for the previous (non-imaging) observations of \lsi\  above 20~keV.  Careful
analysis of {\it RXTE}/HEXTE data allowed \citet{harrison00}  to distinguish
the two sources.  The imaging capabilities of the \intgr\ (several arcminutes
angular resolution) allow, for the first time, to clearly separate the two
sources. Both sources are clearly seen in the  22.1-63.2 keV \intgr\ mosaic of the
 field (Fig.\ref{int_ima}), at 8.1 $\sigma$ (\lsi), and 12.7 $\sigma$ (\qso) detection level. 
\qso\ is several times  brighter than \lsi.  BD +60 536, and HD
+16 429 detected by \xmm\ are too faint  to be observed with \intgr.    

\subsection{Timing analysis}

\begin{figure*}
\begin{center}
\includegraphics[width=14cm,angle=0]{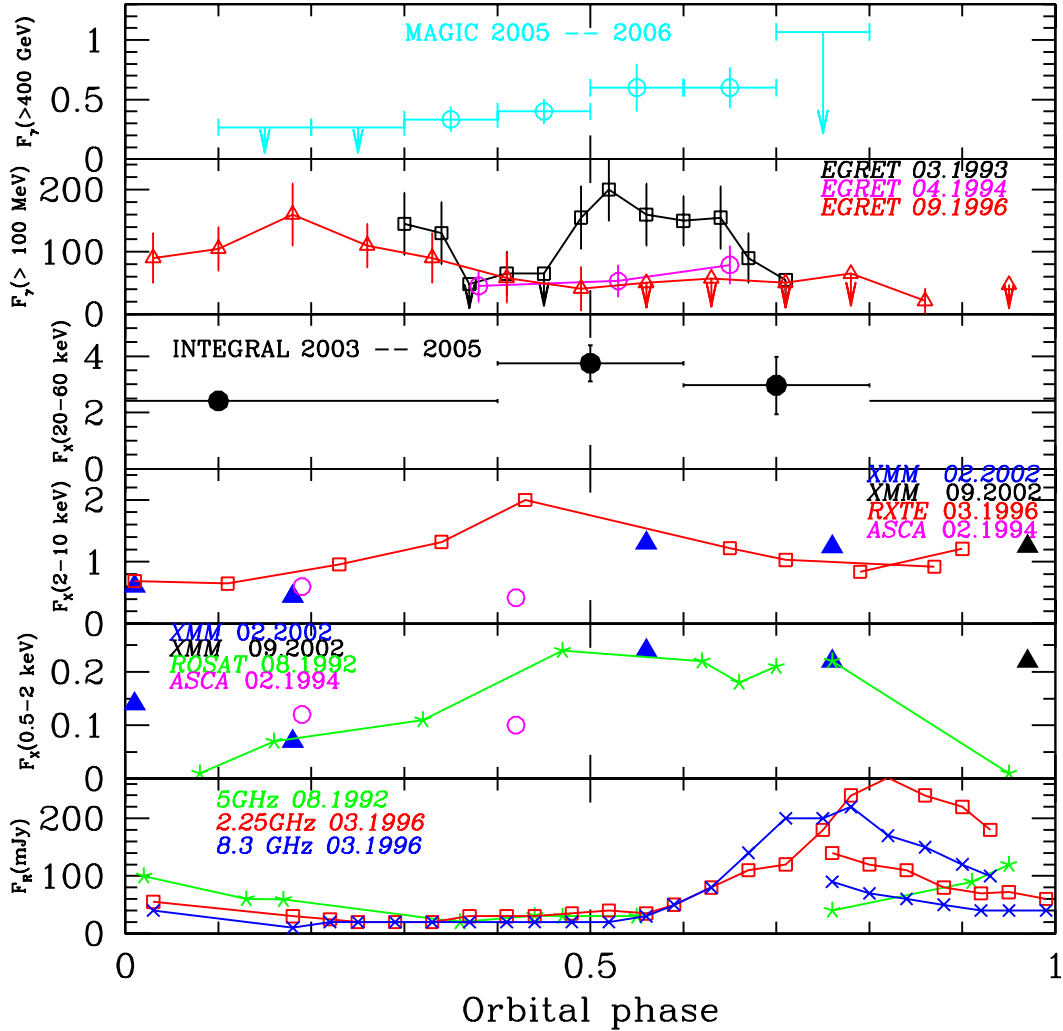}
\end{center}
\caption{Comparison between the TeV (top), GeV gamma-ray orbital lightcurves of
  \lsi\ with the hard X-ray (20-60~keV), X-ray (2-10~keV), soft X-ray 
  (0.5-2~keV)  and radio (bottom) orbital lightcurves. The 0.5 - 2 keV and  2 -
  10 keV X-ray flux is given in $10^{-11}  \mbox{ergs cm}^{-2} \mbox{s}^{-1}$.
  GeV $\gamma$-ray flux ($>$100MeV) is given in the units of $10^{-8}  \mbox{ph
  cm}^{-2} \mbox{s}^{-1}$. TeV flux ($>400$~GeV) is in the units of
  $10^{-11}$~ph~cm$^{-2}$s$^{-1}$). To guide the eye we have connected with
  lines  data from the same orbital cycle. The time is assumed to increase from
  left to right, thus if the observation has started at the end of the orbital
  cycle it is shown with  two lines  of the same color, the line starting at
  larger orbital phase connects data points taken earlier. }
\label{Xradio}
\end{figure*}

Most previous papers on \lsi\ have defined the radio phase according to the
ephemeris  of either \citet{taylor84}, or \citet{gregory99}. In the current
paper we use the  latest ephemeris of the system derived by \citet{gregory02}
($P_{orb}=26.4960$ d, $T_0$=JD 2,443,366.775).  This allows a consistent
treatment  of current and historical observations.   Optical observations
allowed \citet{casares05} to build an orbital solution of the \lsi\ system.
This solution constrains the periastron phase to  $\phi_0=0.23 \pm 0.02$, and
the eccentricity of the system $e=0.72 \pm 0.15$.

Third, forth and fifth (from the top) panels of Fig. \ref{Xradio} show the
dependence  of the 20-60~keV, 2-10~keV and 0.5-2~keV X-ray fluxes  on the
orbital phase. Besides the new \intgr\ and \xmm\ data we show in this Figure
all available archival data from  \asca\ \citep{leahy97}, \rxte\ and
\textit{ROSAT} \citep{harrison00}. 

To allow straightforward comparison with the data in other energy bands we also
show in Fig. \ref{Xradio} the lightcurves of the source in the broad
(radio-to-gamma-ray) energy interval. The uppermost panel shows the highest
energy (TeV) lightcurve reported in \citet{albert06}. The second  panel shows
the EGRET (GeV) data \citep{tavani98}. Finally, the bottom panel of Figure
\ref{Xradio}    shows  radio data simultaneous to \rxte\ and \textit{ROSAT}
observations \citep{harrison00}.

In agreement with previous observations the 2-10~keV flux during \xmm\ 
observations was observed to be  variable with phase.  It was rather low at
zero phase, reached  its   minimum at periastron (near phase 0.2), and became
approximately three times higher at phases larger than 0.5. Comparison of the
X-ray and radio lightcurves of different years shows that the X-ray flux from
the system varies in a more regular way than the radio flux: the X-ray data
points from different orbital cycles lie more-or-less on the same curve, with
the only exception of the second \asca\ point.

\subsection{Spectral Analysis \label{specan}}

\begin{figure}
\includegraphics[width=9cm,angle=0]{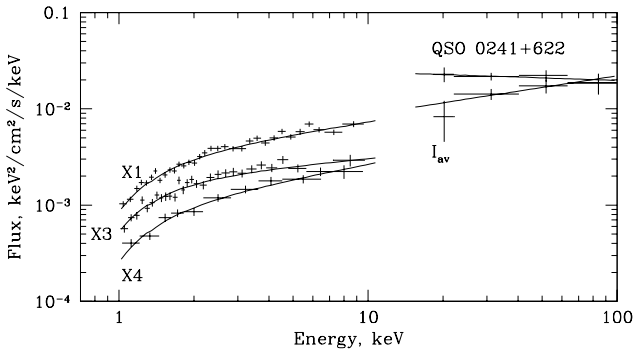}
\vspace {-4cm}
\caption{ \xmm\/ PN \lsi\ spectra from the X1, X3 and  X4 observations. For
\intgr\ the averaged 2003 -- 2004 spectrum is shown. We also show on this plot
the spectrum of the nearby source QSO 0241+622. }
\label{spectry}
\end{figure}

The spectral analysis was done with NASA/GSFC XSPEC software package. In
Fig.~\ref{spectry} the \xmm\ spectra for the X1, X3 and X4  observations , as
well as the average \intgr\ spectrum of the source are shown. We do not show
the spectra of observations X2 and X5 because they are very similar to that of
X1.

For comparison we also give in this Figure the 20-100~keV  spectrum of  \qso\
(the nearby source which was difficult to distinguish from \lsi\ in earlier
non-imaging observations).  The spectrum of \qso\ is well fitted by a powerlaw
with the photon index $\Gamma_{qso}=2.1\pm 0.2$ and 20-100~keV flux $F=(5.5\pm
0.4)\times 10^{-11}$~erg/(cm$^2$s). The source did not show any variability
during the 2-year period of \intgr\ observations.

A simple power law with photoelectric absorption describes the spectrum of
\lsi\ well, with no evidence for any line features. In Table \ref{summary_xmm}
we present the results of the three parameter fits to the \xmm\ data in the 0.5
-- 10 keV energy range. The uncertainties are given at the $1\sigma$
statistical level and do not include systematic uncertainties. The values for
the flux and the powerlaw index were obtained assuming that the hydrogen column
density $N_H$ was the same for all the datasets. The value $N_H=(0.49\pm
0.02)\times 10^{22}$~cm$^{-2}$ was found from the simultaneous  fit of all the
5 \xmm\ datasets. 

\begin{figure}
\begin{center}
\includegraphics[width=\columnwidth,angle=0]{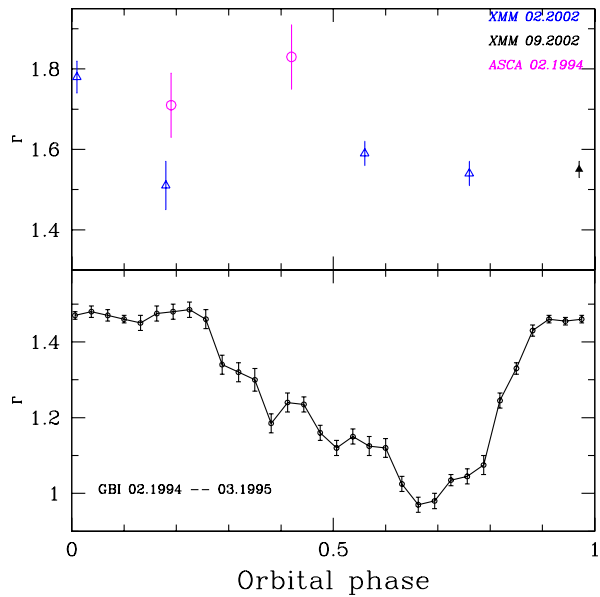}
\end{center}
\caption{Evolution of the  photon index $\Gamma$ as a function of the orbital
phase for \xmm\ and \asca\ (top panel),   and  radio
GBI (bottom panel) observations of the \lsi\ system. 
}
\label{spechist}
\end{figure}

The graphical representation of the evolution of the  spectral parameters along
the orbit is given in Figure~\ref{spechist}  which includes
  soft X-ray from \xmm\ and \asca\ observations
(top panel),  and averaged radio data obtained by \citet{ray97} with  the
National Radio Astronomy Observatory Green Bank Interferometer (GBI) (bottom
panel).

In the \xmm\ energy band the powerlaw photon index is $\Gamma\simeq 1.5$ in all
the observations except for the observation X3 where we find significantly
softer spectrum with $\Gamma=1.78\pm 0.04$. This indicates that the X-ray
spectrum softens during the transition from higher to lower flux state.

\begin{table}
\caption{Spectral parameters for  \xmm\ Observations of \lsi.$^*$}
\label{summary_xmm}
\begin{center}
 \begin{tabular}{l|c|c|c|c}
\hline
Data  & $F$(2-10 keV) &$\Gamma$&$N_H$                &$\chi^2$ (dof) \\
Set   &$10^{-11}$erg s$^{-1}$& &($10^{22}$ cm$^{-2}$)& \\
\hline
X1& 1.30$\pm 0.04$ &1.60$\pm 0.03$&0.49$\pm 0.02$& 474(456) \\ 
X2& 1.24$\pm 0.04$ &1.54$\pm 0.03$&0.49$\pm 0.02$& 440(446) \\
X3& 0.61$\pm 0.03$ &1.78$\pm 0.04$&0.49$\pm 0.02$& 184(188) \\
X4& 0.44$\pm 0.03$ &1.52$\pm 0.06$&0.49$\pm 0.02$& 137(150) \\
X5& 1.25$\pm 0.05$ &1.55$\pm 0.02$&0.49$\pm 0.02$& 488(496) \\
\hline
\end{tabular}
\end{center}
$^*$ Given errors represent 68\% confidence interval 
uncertainties.
\end{table}

\section{Broad band spectrum of the source.}

The rich multiwavelength variability properties of the system, presented in
Figs. \ref{Xradio}, \ref{spectry}, \ref{spechist} suggest that the physical
model of the system which is able to explain all the observed spectral 
variations  should be very restrictive. However, up to now most of the model
building for the source concentrated on explaining the details of either radio
or X-ray or gamma-ray variability, paying little attention to the broad-band
properties of the source. 

As mentioned in the introduction, historically there are two broad classes of
physical models of the source activity. The first one  assumes that the
activity  is powered by accretion onto the compact object (either neutron star
or a black hole) \citep{taylor84}. Accreting X-ray binaries  with a Be star
companion constitute the majority of the Galactic high-mass X-ray binaries
\citep{ziolkowski02}. Most of the Be star X-ray binaries contain an accreting
neutron star as the compact object and most of them are transient sources. The
spectra of these sources are characterized by the presence of an exponential
cut-off  in the hard X-ray band at energies 10-60~keV
\citep{white83,filippova05}. Our analysis of the \intgr\ data shows that in the
case of \lsi\ no high-energy cut-off is found at the energies below 100~keV, the fact that does not fit well 
into  the "conventional" accretion scenario.  It is
possible that in the case of \lsi\ emission from the accretion disk is "masked"
by the emission from the jet, which can have a simple powerlaw spectrum. If
accretion flow and jet would give comparable contributions in the hard X-ray
band, one would still expect to observe a spectral feature between 10 and
100~keV which would result from the disappearance of the accretion component at
higher energies. The non-observation of such feature in our data indicates that
the luminosity of the accretion flow in \lsi\ is much lower than the observed
X-ray luminosity, $L_{\rm accretion}\ll  10^{34}$~erg/s. Since, by assumption,
the source activity (including the jet) is powered by accretion,  the large
apparent luminosity of the jet ($L_{jet}\sim 10^{35}$~erg/s, of the order of
the gamma-ray luminosity)  can be explained by a  large beaming factor of the
jet, $\delta\simeq 2$. 

Radio observations suggest that the jet is moderately relativistic, with the
speed $\beta\sim 0.4\div 0.6$. It is presumably precessing and the  angle with
respect to the line of sight varies from  $\theta\simeq 0^\circ$
\citep{massi01} to $\theta\simeq 78^\circ$ \citep{massi04}. This implies that
the Doppler factor  of the jet,
$\delta=\sqrt{1-\beta^2}/(1\pm\beta\cos\theta)$, varies in the range
$1<\delta<2$ along the orbit. If the high  gamma-ray luminosity of the source
during a fraction of the orbit is explained by the large Doppler  factor of the
jet  ($L_{jet}\sim \delta^4$), one would expect that the  jet contribution
drops during the orbital phase with the low Doppler factor.  In this case the
"accretion" contribution would show up in the low X-ray flux state. However,
our data show that the spectrum of the source is well fit by a featureless
powerlaw with the same photon index both in the high and low X-ray states. 
It is not impossible that the accretion component is, similarly to jet component, 
modelled by a powerlaw, but in this case the coincidence of the photon indexes in the high and low flux states
implies an artificial "fine tuning" of parameters.
Although we find little support for the accretion model in our data, 
further multiwavelength observations are necessary to better constrain the
Doppler factor and the hard X-ray spectral properties of the source low
state.   

\begin{figure*}
\includegraphics[width=1.9\columnwidth,angle=0]{LSI_new.epsi}
\caption{Broad band spectrum of \lsi. Radio data points are taken from 
\citet{strickman}. {\it CGRO} data points are from \citet{tavani96,tavani98}. MAGIC data points 
are from \citet{albert06}. The solid (dashed) line shows the model fit within the synchrotron-inverse Compton model
for the high (low) flux state of the source. The dotted line shows possible contribution from the proton proton interactions.
The values of parameters of the model
fits are cited in the text.}
\label{model}
\end{figure*}

The timing and spectral properties in the keV-MeV energy band  favor the second
class of models, which assume that \ the system activity is powered by the
rotation energy of a relatively young radio pulsar \citep{maraschi81}. At least
one more Be star X-ray binary which is powered in this way is known, namely,
PSR B1259-63. In fact, the spectral properties of PSR B1259-53 system are very
similar to the \lsi: strong radio flares and non-thermal X-ray and gamma-ray
emission are detected during the periods of pulsar passing through the disk of
the Be star \citep{johnston05,chernyakova06,aharonian05}. In the model of
\citet{maraschi81} the radio emission is the synchrotron emission from the
pulsar wind electrons while the X-rays are produced via the inverse Compton
scattering of the soft photons from the Be star by the same electrons. Roughly
the same conclusion was deduced from the recent \xmm\ observations for  PSR
B1259-63 \citep{chernyakova06}. The main differences between the two systems
are that (a) the pulsar itself was never detected in the case of \lsi\ and (b)
in the case of PSR B1259-63 no AU-scale jets were seen (one should note that
PSR B1259-63 is in the Southern hemisphere where no VLBI observations are
available).

In the case of the "young pulsar" scenario, radio, X-ray and gamma-ray emission
come from the region where the pulsar wind interacts with the wind of Be star.
The high-energy particles responsible for the synchrotron and inverse Compton
emission can either originate from the relativistic pulsar wind or be
accelerated in the shock formed at the interface of the stellar and pulsar
wind.

 In the inverse Compton model of X-ray emission the soft photons  upscattered
to the X-ray energies  are the UV photons emitted from the Be star. If the
distance from the X-ray emission region to the star is $R$, the energy density
of UV photons at the location of the pulsar can be estimated as
\begin{equation} 
\label{rad} 
U_{rad}=\frac{L_*}{4\pi R^2c}=2\times
10^2\left[\frac{L_*}{10^{38}\mbox{ erg/s}}\right]\left[\frac{10^{12}\mbox{
cm}}{R}\right]^2\mbox{ erg/cm}^3 
\end{equation} 
where $L_*$ is the luminosity of the Be star. Relativistic electrons upscatter 
the UV photons  up to  energies 
\begin{equation} 
\epsilon_{IC}\simeq
4\left[\frac{T}{2\times 10^4\mbox{ K}}\right] \left[\frac{E_e}{10\mbox{
MeV}}\right]^2\mbox{ keV} 
\end{equation} 
where $T$ is the temperature of the Be star. Apart from producing inverse
Compton emission in the X-ray band, the 1-10~MeV electrons emit radio 
synchrotron radiation. The typical frequency of the synchrotron photons is 
\begin{equation} 
\epsilon_s\simeq
7\left[\frac{B}{1\mbox{ G}}\right]\left[\frac{E_e}{10\mbox{
MeV}}\right]^2\mbox{ GHz} 
\end{equation} 
where $B$ is the magnetic field strength in the emission region.  In the
synchrotron -- inverse Compton model the ratio of the radio and X-ray
luminosities of the system, $L_{\rm radio}/L_{\rm X-ray}$,   is equal to the
ratio of the energy densities of magnetic field,  
\begin{equation} 
U_B\simeq 4\times
10^{-2}\left[\frac{B}{1\mbox{ G}}\right]^2\mbox{ erg/cm}^3 
\end{equation} 
to the energy density of radiation (\ref{rad}): 
\begin{equation} 
\label{uburad}
\frac{L_{\rm radio}}{L_{\rm X-ray}}=\frac{U_B}{U_{rad}}\simeq 2\times 10^{-4}
\left[\frac{B}{1\mbox{ G}}\right]^2\left[\frac{R}{10^{12}\mbox{ cm}}\right]^2
\left[\frac{10^{38}\mbox{ erg/s}}{L_*}\right] 
\end{equation} 
The last relation imposes a restriction on the possible values of magnetic
field in the emission region. Taking into account that the distance from Be
star to the compact companion is constrained by the binary orbit parameters
roughly in the limits $10^{12}\mbox{ cm}\le R\le 10^{13}\mbox{ cm}$ and taking
the observed ratio $L_{\rm radio}/L_{\rm X-ray}\sim 10^{-3}$, the magnetic
field is constrained to be  
\begin{equation} 
B\le (several)\mbox{ G} 
\end{equation} 

An example of synchrotron -- inverse Compton model fit of the broad band (radio
to gamma-ray) spectrum of the source is shown in Fig. \ref{model}. The model
parameters used for the fits of the broad band spectrum of the system in high
(low) flux state in Fig. \ref{model} are $B=0.35$~G ($B=0.25$~G), the electron
spectrum described by a broken powerlaw with the spectral index $\Gamma_e=2$ 
below the break energy $E_{br}=2\times 10^8$~MeV ($E_{br}=10^8$~eV) and
$\Gamma_e=3.5$ ($\Gamma_e=4$) above the break energy.  In the case of low radio
state the break is at a bit lower energy and the electron spectrum is softer
above the break energy,  to fit the radio spectrum of the source which clearly
exhibits a cut-off at  $\sim 10$~GHz frequencies. One can see that in this case
the inverse Compton spectrum softens at the energies $E\sim 100$~keV.

It is interesting to note that even if the electron spectrum extends to
energies higher than 10~GeV,  one still expects to observe a softening of the
gamma-ray spectrum at 10~GeV energy because above this energy the inverse
Compton scattering proceeds in the Klein-Nishina regime and if the electron
spectrum is a powerlaw with the spectral index $\Gamma_e$, the inverse Compton
spectrum below 10~GeV is a powerlaw with the photon index $(\Gamma_e+1)/2$
($dN_{\gamma}/dE\sim E^{-(\Gamma_e+1)/2}$) while above 10~GeV the spectrum
softend to  roughly $dN_{\gamma}/dE\sim E^{-(\Gamma_e+1)}\ln E$. The spectrum
above 400~GeV measured by  MAGIC telescope \citep{albert06}  is indeed softer
than the EGRET spectrum in the  GeV band. This can be an indication of
transition to the Klein-Nishina regime. However, the observed softening is less
than expected in this case and it is possible that emission above 100~GeV  is
dominated by another mechanism (e.g. by  pion decay gamma-rays,  as suggested 
by \citet{romero05}). We come back to this possibility in Section 5.

The X-ray luminosity of the system is several orders of magnitude higher than
radio luminosity. This enables to conclude that the inverse Compton energy loss
always dominates over the synchrotron energy loss.  Indeed, the inverse Compton
cooling time is
\begin{equation}
\label{tic}
t_{IC}\simeq 10^4\left[\frac{10^2\mbox{ erg/cm}^3}{U_{rad}}\right]\left[\frac{1\mbox{
keV}}{\epsilon_{IC}}\right]^{1/2}\mbox{ s}
\end{equation}
while the synchrotron cooling time is
\begin{equation}
t_{synch}\simeq 10^6\left[\frac{1\mbox{ G}}{B}\right]^{3/2}\left[
\frac{1\mbox{ GHz}}{\epsilon_s}\right]^{1/2}\mbox{s}
\end{equation}

Factor of 10 variations of the inverse Compton (X-ray) luminosity of the source
can be produced  if  the typical distance from the emission region to Be star
varies by a factor of 3. This is smaller than the variations of the distance
from the compact object to Be star, estimated from the eccentricity of the
binary orbit. But, in fact, in the models of interaction of the pulsar wind
with the stellar wind from the companion star the bulk of emission from the
system is supposed to come from the shock region at the contact surface of the
two winds, which is situated in between the star and the compact object and
possibly extends to the distances larger than the binary separation. 

In order to explain the observed spectral properties of the source,  one has to
compare the inverse Compton cooling time to the escape time from the emission
region. Several velocity scales are present at the interface of stellar and
pulsar winds.  If the pulsar wind mixes with the stellar wind, the escape of
the high-energy particles could be slowed down to the  velocity of the "slow"
equatorial wind from Be star,  $v_{disk}\sim 10^6\div 10^7$~cm/s. The same
estimate of the escape velocity  is obtained if the high-energy particles are
stellar wind particles accelerated in the shock at the interface with the
pulsar wind. To the contrary, if the two winds flow along the contact surface
without significant mixing, the escape velocity of the pulsar wind particles is
comparable to the speed of light, $v_{pulsar}\sim 10^{10}$~cm/s.
Correspondingly, two different scenarios are possible: that of a "slow escape",
with the escape time $t_{esc}\sim R/v_{disk}\sim 10^6$~s and that of a "fast
escape", with $t_{esc}\sim R/v_{pulsar}\sim 10^3$~s. 

In the case of "slow escape", the inverse Compton emission should be  produced
by the cooled electron population.  For a powerlaw distribution of electrons,
the steady state spectrum has a photon index $\Gamma_e$ which is softer than
the  spectral index $\Gamma_{e,\rm inj}$ of injection spectrum, 
$\Gamma_e=\Gamma_{e,\rm inj}+1$. The observed photon index in X-ray band is
$\Gamma_{ph}=(\Gamma_{e}+1)/2\simeq 1.5$. This implies that  $\Gamma_{e,\rm
inj}+1\simeq 2$, which means that the injection spectrum should be very hard,
$\Gamma_{e,\rm inj}\simeq 1$. Such spectrum hardly can be produced by shock
acceleration at the contact surface of pulsar and stellar winds. Thus, in the
"slow escape" scenario the only possibility is that  electrons are initially
injected only at high energies (that is,  they originate from the cold pulsar
wind with large bulk Lorentz factor) and the inverse Compton cooling would
leads to the formation of "universal" cooling spectrum with $\Gamma_e=2$,
regardless of the details of the injection spectrum at high energies. 

In the case of "fast escape" the high-energy electrons also originate in the
pulsar wind (by assumption). In this case the inverse Compton cooling time at
1-100~keV energies is comparable to the escape time and, in the case of
powerlaw injection spectrum,  one expects to observe a "cooling break"
(softening by $\Delta\Gamma_{ph}\simeq 0.5$) at the energy at which the escape
time is equal to the cooling time. From the model fit to the broad-band
spectrum, shown in Fig \ref{model}, one can see that a break could be situated
at the energies above MeV (electron energies $E_{e,br}\sim 100$~MeV) which is,
in principle, compatible with the "fast escape" scenario. In fact, the "fast
escape" scenario is not so different from the "slow escape" one, because  the
spectrum of the high-energy electrons from the pulsar wind with large Lorentz
factor has a low-energy cut-off. Escaping electrons cool down  only to the
characteristic energy $E_{e,br}\sim 100$~MeV in the direct vicinity of the
pulsar and Be star. As a result, 100~MeV electrons are injected in a larger
region around the binary system and continue to cool (at a longer time scale)
forming a characteristic $E^{-2}$ powerlaw spectrum below 100~MeV.

The observations presented in the previous sections clearly reveal the
softening of the spectrum  during the decrease of the X-ray luminosity.  The
softening of the spectrum can be explained assuming that the injection of
electrons at high energies drops and  higher energy electrons (emitting in the
hard X-ray band) cool faster than the low energy ones (emitting in the soft
X-ray  band). Note that this explanation implies that the size of the region
from which the bulk of X-ray emission comes  is at least $R\sim 10^{13}$~cm,
otherwise the inverse Compton cooling time (\ref{tic})  would be too short to
explain the softening on day time scale.

Although radio and X-ray emission are produced by the same population
of electrons, they are not necessarily produced in the same spatial region.
The strongest synchrotron emission is produced in the 
region with higher magnetic field which is situated in the direct vicinity of 
the pulsar, while the strongest inverse Compton emission is produced in the
region with highest soft photon background density, which is situated close to
the companion star. This means that, in general, one does not expect a strong correlation 
of the X-ray and radio lightcurves. Taking into account the fact that 10-100~MeV electrons 
can propagate over a significant distance during the inverse Compton cooling time
(\ref{tic}), the detailed modelling of the X-ray and radio lightcurves requires 
numerical simulations, which would take into account (somewhat uncertain and time-variable) 
geometrical  properties of the system, such as, e.g., the geometry and density profile of the equatorial disk 
of Be star and its inclination to  the orbital plane. We leave this for the future work.

\section{Physical model of the source.}

The above modeling of the broad-band spectrum of the system shows that the
physical model of the source (which we assume to be based on the "young pulsar
powered source" model, see above) should explain several key features derived
from the analysis of radio, X-ray and gamma-ray data. 

In particular, we have seen that the X-ray spectrum and spectral variability
are satisfactory explained if the relativistic electrons responsible for the
X-ray emission are initially injected at  high energies and subsequently cool
down forming the characteristic $E^{-2}$ cooling spectrum. The physical model
should provide an explanation for the electron injection only at high
energies.  The analysis of the radio and X-ray variability of the system shows
that the injection of  high-energy electrons should be variable.  

Injection from  high energies can be explained if the electrons responsible for
the X-ray emission originate from the cold pulsar wind with bulk Lorentz 
factor $\ge (several)\times 10^2$. In this case all electrons have initial
energies larger  than 100~MeV and the electron spectrum below 100~MeV formed in
the process of inverse Compton cooling has the spectral index $\Gamma_e\simeq
2$. However, an immediate difficulty with such simple injection model is that
the injection rate is not expected to vary with time, contrary to what is
observed.

The variable injection rate could be provided if relativistic protons are
either present in the pulsar wind or accelerated in the shock at the contact
surface of pulsar and stellar wind. Such relativistic protons could interact
with the low energy protons from the disk and produce injection of electrons
at  energies above $\sim 100$~MeV via production and subsequent decays of
charged pions. The pion production cross section is $\sigma_{pp}\sim
10^{-26}$~cm$^2$. The life time  of high-energy protons in the disk with the
density $n\sim 10^{10}$~cm$^{-3}$ is $t_{pp}=1/(\sigma_{pp}nc)\simeq 10^{6}$~s,
comparable with the escape time from the disk. This means that (a) protons can
efficiently transmit their energy to the products of pion decay (gamma-rays,
neutrinos, electrons) during their propagation through the disk, (b)
proton-proton interactions can provide the source for injection of 10-100~MeV
electrons over the large AU-size region, from which the bulk of X-ray emission
supposedly comes and (c) variable injection rate of high-energy electrons is
explained by the variations in the density of the stellar wind protons along
the pulsar orbit. 

A straightforward consequence of high-energy proton interactions in the Be star
disk is the appearance of additional component in the high-energy $\gamma$-ray
spectrum, resulting from the two photon decays of neutral pions. The spectrum 
of the pion decay gamma-rays in the GeV-TeV energy band has the same spectral
index as the spectrum of the high-energy protons. In  Fig. \ref{model} we show
possible contribution to the $\gamma$-ray spectrum of the source which can be
produced in result of proton-proton interactions.

If the system activity is powered by a young pulsar, the pulsar  should be also responsible 
for the formation of the 100 AU scale jet  in the system. The fact that rotation powered
 pulsars can produce jets  is beyond doubt, since such jets are observed in the pulsar wind  
 nebulae of e.g. Crab and Vela pulsars. In the case of the pulsar  interacting with the wind
 of companion star a formation of even more  complicated extended emission features which can 
 immitate the jets is  possible \citep{dubus06}. However, no "immitation" of the jet is 
  really needed, because the mechanism of formation of the jet in the  system of a young pulsar
   interacting with a stellar wind can be the  same as in the system of a young pulsar interacting
    with interstellar  medium. The naive qualitative estimate shows that anisotropic power  
    injection into the stellar wind should, in general, lead to the  formation of an outflow from 
    the system with parameters close to the  expected physical parameters of the jet. 
 Indeed,   the total energy which is injected during the escape
time scale is $E_{kin}\sim P t_{esc} \sim 10^{41}$~erg. Such kinetic energy
transmitted to a volume of the size of about the thickness of the disk, $H\sim
10^{12}$~cm with the density $n\sim 10^{10}$~cm$^{-3}$ (total mass is $M\sim
m_pnH^3\sim  10^{22}$~g pushes the matter in the volume to move with the speed 
\begin{equation} 
v_{bulk}\sim \sqrt{\frac{E_{kin}}{M}}\sim
0.1c\left[\frac{P}{10^{35}\mbox{ erg/s}}\right]\left[\frac{t_{esc}}{10^6\mbox{
s}}\right]\left[\frac{10^{10}\mbox{ cm}^{-3}}{n}\right]\left[
\frac{10^{12}\mbox{ cm}}{H}\right] 
\end{equation} 
This velocity is of the order of the velocity of the fast polar wind from Be
star  and is much larger than the typical velocity inside the disk,
$v_{disk}\sim 10^6$~cm/s.  Qualitatively, it is possible that (anisotropic) 
injection of large kinetic energy of the pulsar wind into the disk is
responsible for the formation of a jet-like outflow (or "expanding bubbles"
implied in the models of radio flares) from the disk in which relativistic
protons and electrons from the pulsar wind are mixed with the low-energy
protons from the disk. The proton-proton interactions resulting in production
of pions which subsequently decay and inject high-energy electrons can continue
also in the outflow thus providing a mechanism of injection of high-energy
electrons at larger scales up to the scale of the $\sim 100$~AU jet observed in
the system.

\section{Summary.}

In this paper we have presented the \intgr\ and \xmm\ observations of  \lsi. We
have found that the overall spectrum of the system in 0.5-100~keV band is well
fit by a featureless powerlaw with the photon index $\Gamma_{ph}\simeq 1.5$
both in the high and low flux states. The 0.5-100~keV powerlaw spectrum matches
smoothly the higher-energy spectrum of the source in 100~keV -- GeV band. The
powerlaw with the same photon index seems to continue without a cut-off up to 
1-10~MeV energies. Non-observation of a cut-off or a break in the spectrum at
10-100~keV energies, typical for the accreting neutron stars and black holes,
implies that the accretion in the source proceeds at a very low (if not zero)
rate.  This favors the scenario in which the compact object is a
rotation-powered pulsar.  We have discovered that the spectrum is hard in both
high and low X-ray states, but  softens during the transition from the high
flux to the low flux state.  The spectral characteristics of the system favor
the model in which electrons are initially  injected at high energies and
subsequently cool to lower energies forming the characteristic "cooling"
spectrum with the spectral index $\Gamma_e\simeq 2$. We have proposed a model
in which  cold relativistic pulsar wind (electron and/or proton loaded) with
the bulk Lorentz factor $\Gamma\sim 10^2\div 10^3$ provides a  source of X-ray
emitting electrons. Such relativistic wind penetrating into the Be star disk
can also be responsible for the formation of jet-like outflow from the system. 

\section{Acknowledgments}

The authors acknowledge useful discussions with  F.~Aharonian, I.~Kreykenbohm,
J.~Rodriguez, M.~T\"urler and V.Bosch-Ramon.

 \label{lastpage}

\end{document}